# A wideband amplifying and filtering reconfigurable intelligent surface for wireless relay

*Lijie Wu* [† 1], *Qun Yan Zhou* [† 1], *Jun Yan Dai* [† * 1,2,3], *Siran Wang* [† 1,4], *Junwei Zhang* [1], *Zhen Jie Qi* [1], *Hanqing Yang* [1], *Ruizhe Jiang* [1], *Zheng Xing Wang* [1], *Huidong Li* [1], *Zhen Zhang* [6], *Jiang Luo* [7], *Qiang Cheng* [* 1,2,3], *and Tie Jun Cui* [* 1,2,5]

[1] State Key Laboratory of Millimeter Waves, Southeast University, Nanjing 210096, China

[2] Institute of Electromagnetic Space, Southeast University, Nanjing 210096, China

[3] Frontiers Science Center for Mobile Information Communication and Security, Southeast University, Nanjing 210096, China

[4] State Key Laboratory of Terahertz and Millimeter Waves, City University of Hong Kong, Hong Kong, China

[5] Suzhou Laboratory, Suzhou 215000, China

[6] School of Electronics and Communication Engineering, Guangzhou University, Guangzhou 510006, China

[7] School of Electronics and Information, Hangzhou Dianzi University, Hangzhou, 310018, China

[†]Equally contributed to this work

[*]E-mail: junyand@seu.edu.cn, qiangcheng@seu.edu.cn, and tjcui@seu.edu.cn

## Abstract

Programmable metasurfaces have garnered significant attention due to their exceptional ability to manipulate electromagnetic (EM) waves in real time, leading to the emergence of a prominent area in wireless communication, namely reconfigurable intelligent surfaces (RISs), to control the signal propagation and coverage. However, the existing RISs usually suffer from limited operating distance and band interference, which hinder their practical applications in wireless relay and communication systems. To overcome the limitations, we propose an amplifying and filtering RIS (AF-RIS) to enhance the in-band signal energy and filter the out-of-band signal of the incident EM waves, ensuring the miniaturization of the RIS array and enabling its anti-interference ability. In addition, each AF-RIS element is equipped with a 2-bit phase control capability, further endowing the entire array with great beamforming performance. An elaborately designed 4×8 AF-RIS array is presented by integrating the power dividing and combining networks, which substantially reduces the number of amplifiers and filters, thereby reducing the hardware costs and power consumption. Experimental results showcase the powerful capabilities of AF-RIS in beam-steering, frequency selectivity, and signal amplification. Therefore, the proposed AF-RIS holds significant promise for critical applications in wireless relay systems by offering an efficient solution to improve frequency selectivity, enhance signal coverage, and reduce hardware size.

**Introduction**

In recent years, the fifth-generation (5G) networks have been deployed in over ten countries, prompting researchers to gradually shift their focus towards the sixth-generation (6G) networks[1–6]. Owing to the proliferation of ubiquitous wireless connectivity driven by Internet of Things (IoT) technology, there is a substantial demand for higher system capacity in the 6G networks[7–10]. Meanwhile, the increasing number of access devices has led to urgent challenges in enhancing signal strength, ensuring communication quality, and addressing energy consumption. In the potential 6G technology solutions, the ultra massive multiple input multiple output (MIMO) system can utilize a high-gain antenna array to enhance the signal strength within the cell[11–15]. However, due to the presence of buildings, plants, and other obstructions, there are still signal blind zones in the entire cell area, affecting the smooth experience of users. To solve the problem, repeaters, relays, and other wireless devices are introduced to enhance the signal strength in blind zones to improve the communication quality[16,17]. However, the traditional relay devices are somewhat inadequate for meeting the requirements of 6G communication systems, owing to their high costs, high energy consumption, and complex architectures[18,19]. The repeaters may have a lower cost, but they cannot realize the beamforming. Sometimes the omnidirectional coverage of repeaters instead increases intra-cell interference, resulting in complex communication network planning.

To solve these issues, programmable metasurfaces, also known as reconfigurable intelligent surfaces (RISs) in the communication community, have gained widespread attention due to their distinctive attributes of low costs, low energy consumption, and simple architectures. Typically, A programmable metasurface or RIS comprises a large number of meticulously designed periodic structures loaded with tunable elements such as positive-intrinsic-negative (PIN) diodes and varactor diodes. By dynamically controlling the tunable elements through a micro-control unit (MCU) or field-programmable gate array (FPGA), the programmable metasurface can achieve real-time controls of all fundamental characteristics of electromagnetic (EM) waves, including amplitude, phase, polarization, frequency, and wavevector[20–28].

Owing to its excellent EM manipulation performance, the programmable metasurface has been widely explored to build a flexibly controlled EM environment, enabling a groundbreaking capability to reconfigure the wireless channels as RIS[29–32]. The programmable metasurface and RIS serve a crucial role in the realm of wireless communications, such as simplified transmitter architectures[33–40] and innovative wireless relays[41–44]. In particular, RISs can be strategically deployed in both outdoor and indoor environments and integrated with existing systems for joint optimization of wireless channels. The integration aims to reshape the wireless environment, thereby enhancing the signal-to-noise ratio (SNR) and expanding the signal coverage range[45–47]. Hence RIS presents a promising solution for upcoming 6G communication systems, providing a simple and highly efficient wireless relaying capability, and has been selected as one of the top 10 emerging technologies of 2024[48].

Nevertheless, the effectiveness of this technology is sometimes hindered by the additional path loss attenuation in the relay links, resulting from a significant reduction of scattering energy at the material interface due to the non-negligible loss of the tunable elements in RISs. Hence, larger RIS arrays are often required to increase the array gain to ensure adequate signal strength, leading to higher hardware costs. To mitigate this problem, programmable metasurfaces with EM-wave amplification functions or amplifying RISs (A-RISs) have emerged as the potential solution. As indicated in Refs. 49–52, the introduction of A-RIS to compensate for the path loss yields substantial improvements in energy efficiency, sum-rate gain, and signal coverage in metasurface-aided systems. However, these papers were predominantly focused on proposing the theoretical models of A-RIS and conducting numerical analyses of system performance, lacking practical and detailed RIS design and verifications. Only in Ref. 53, the signal model of A-RISs was validated by experiments on a constructed A-RIS element. Nevertheless, the intricate design and large physical size of the A-RIS element significantly restrict its feasibility for practical implementation in real-world systems.

Typically, the amplifying programmable metasurface and A-RIS are intricately constructed using design principles of amplified reflectarray antennas[54–56]. Simple or complicated amplifier circuits are integrated into the metasurface to realize various remarkable functions such as non-reciprocal scattering[57–59], reflection enhancement[60–63], spatial frequency multiplication[64], simultaneous wireless information and power transfers[65], and construction of programmable diffractive neural networks[66]. However, most of the aforementioned amplifying programmable metasurfaces and A-RISs work in a narrow band or lack phase tuning capability. They are typically equipped with an amplifier in each unit, resulting in higher hardware cost and power consumption.

Besides those problems, the potential spectrum pollution should also be considered. As analyzed in [67,68], existing RISs not only manipulate signals within the desired frequency range but also impact signals outside of this range, leading to significant network interference issues or potential security risks. The mentioned A-RISs even enhance these undesired signals, which may cause even more severe deterioration of the whole communication system. This lack of frequency selectivity must be addressed earnestly when we introduce the RISs or A-RISs into the real wireless environment as the spectrum is becoming more congested. To address this issue, the concept of filtering RISs (F-RISs) has emerged as a promising remedy. As indicated in Refs. 69,70, the proposed F-RISs show a superior frequency-selecting feature and great beam-steering ability. However, the mentioned F-RISs face unavoidable loss in the passband due to the lack of signal amplification.

In summary, both A-RISs and F-RISs have typical limitations, which significantly curtail their practical application in the context of the 6G wireless relay systems. To solve these problems, here we combine the merits of A-RISs and F-RISs and propose the concept of amplifying and filtering RIS (AF-RIS). In this article, an AF-RIS is elaborately designed to achieve substantial in-band energy enhancement and out-of-band signal rejection for the incident EM waves. In addition, 2-bit phase tuning is also realized to achieve the dynamic beam-steering ability. Furthermore, through the introduction of a power combining and dividing network, a 4×8 AF-RIS array is constructed with fewer amplifiers and filters, which greatly reduces the power consumption and hardware cost. The simulated and measured results are in good agreement, affirming the dynamic energy amplification, frequency selectivity, and beamforming capability of the reflected EM wave. A simple and low-cost AF-RIS-based wireless relay system is further established to validate the reliability of the AF-RIS array. Leveraging the advantages of the proposed AF-RIS array mentioned above, it is anticipated to have significant implications in next-generation communication technologies and wireless systems.

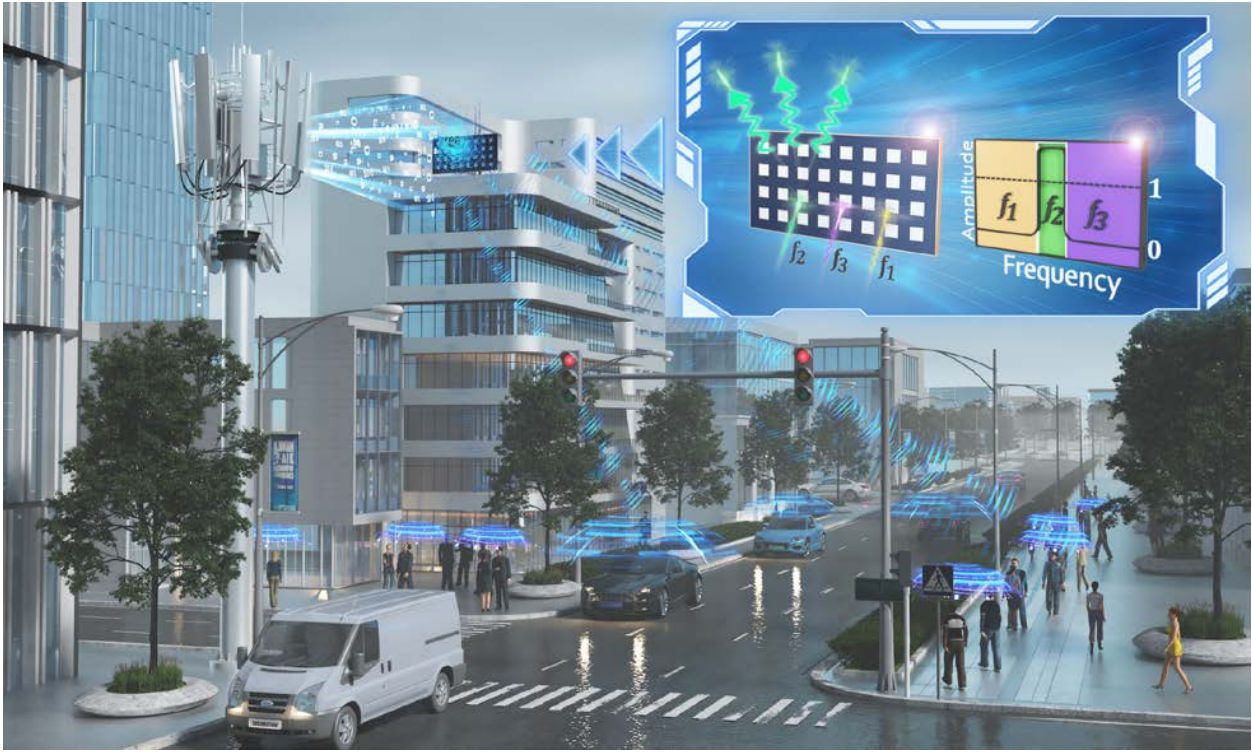

Figure 1. Conceptual diagram of the AF-RIS arrays acting as novel relays in wireless communications.

## Results

Figure shows the conceptual diagram of the proposed AF-RIS array acting as a novel wireless relay. The AF-RIS array can implement EM beamforming, frequency selection, and energy amplification. When installed on the building, it can receive and amplify signals of desired frequency from the base station (BS) while filtering the signal of undesired frequency, and subsequently transmit them to the user ending (UE). This capability significantly extends the coverage range of wireless networks and enhances signal quality, particularly in scenarios where obstacles such as trees along roads, buildings, and vehicles severely obstruct the links between the BS and the UEs.

**Design of AF-RIS**

As depicted in Fig. 2a, b, the AF-RIS array consists of four AF-RIS subarrays, and each subarray is comprised of four main parts: the top patches, the middle slot plane, the microstrip network, and the bottom ground. For a better illustration of the working mechanism, a detailed view of the slot plane and microstrip network is presented in Fig. 2c. Specifically, the impinging spatial waves are received by the patches, converted to guided waves through the slots, and injected into the microstrip network. Firstly, the signals are collected by the power-combining network. Then, they are transmitted to the filtering and amplifying circuit, which provides great out-of-band rejection and in-band energy enhancement. After this, the signals are distributed to each AF-RIS element for phase tuning. Finally, the signals are coupled and converted to the spatial waves through the different slots for orthogonal reradiation. More details on the phase tuning, the power combining and dividing network, the filtering and amplifying circuit, and the design of the AF-RIS element can be found in Supplementary Note 1-3. Based on the meticulous design of the geometrical structure, the AF-RIS exhibits three main advantages, listed as follows for better illustration.

**Phase-tuning and beam-steering capability**

The phase-tuning ability of AF-RIS is realized by the integrated 0°-90° phase-shifter and 0°-180° switch in each element. By switching the ON or OFF states of PIN diodes, signals can traverse different microstrip paths to achieve four reconfigurable states with a 90° interval. To assess the phase-tuning and beam-forming ability of the AF-RIS, full-wave simulations are carried out with CST MWS and a subarray of AF-RIS is analyzed here. In the simulation of phase tuning, each AF-RIS element of the subarray shares the same phase state and as we change the phase state (S0, S1, S2, S3) of all the elements simultaneously, different phase responses are obtained and shown in Fig. 2d. It is observed that stable 90° phase differences are exhibited between the curves, which enables a good 2-bit phase coding characteristic within the passband from 2.8 GHz to 3.2 GHz. By independently controlling the phase state of each element, the beam-steering on the *yoz* plane is enabled. In the passband, the eight AF-RIS elements of the subarray are encoded with seven different coding sequences for examples. The scattering patterns of the reflected waves at 3 GHz are plotted in Fig. 2e, showing that the AF-RIS subarray can reflect the beams to the direction with the angles of ±30°, ±20°, ±10°, and 0°. Besides, to visually demonstrate the beam-steering ability, Figure 2i displays the intensity distribution of the electric field (E-field) on the *yoz*-plane at 3 GHz as the AF-RIS array redirects the incident beams to different angles, which are determined by the coding sequences, i.e., the phase distributions on AF-RIS.

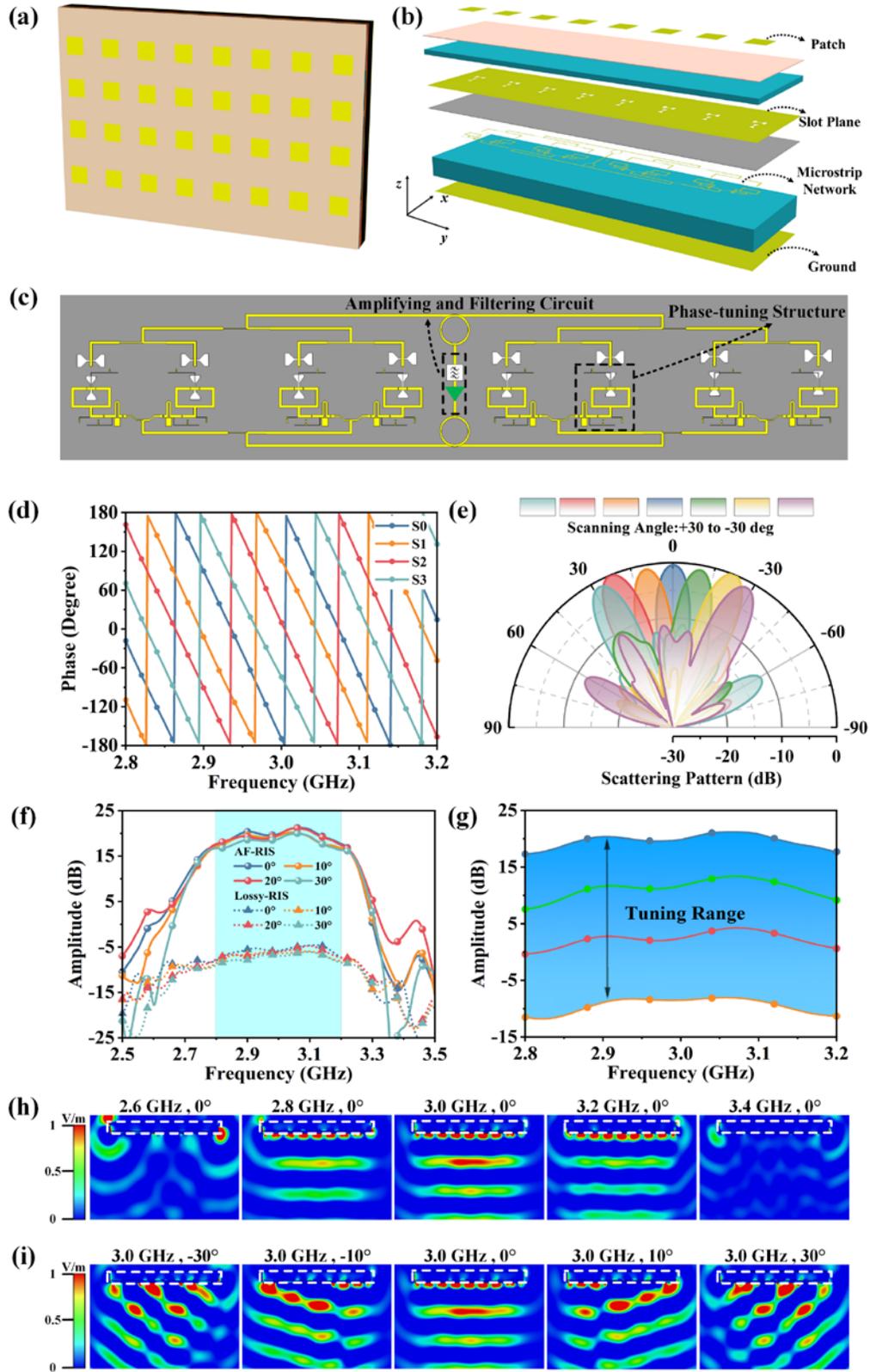

Figure 2. (a) Front view of the 4*8 AF-RIS array. (b) Exploded view of the proposed 1*8 AF-RIS subarray. (c) Top view of the slot plane and the microstrip network of the AF-RIS subarray. (d) The simulated normalized

scattering patterns of the passive AF-RIS array. (e) Active and passive reflection amplitude of the AF-RIS array under normal incidence with the reflected beams steering to four different angles. (f) The E-field intensity distributions on the *yoz*-plane. (g) The E-field intensity distributions on the *yoz*-plane while the reflected beams point to different directions.

**Amplifying and filtering properties**

AF-RIS also exhibits great in-band energy enhancement and out-of-band rejection performance, owing to the integrated amplifying and filtering circuit. For comparison, the lossy RIS is considered here. The structure of the lossy RIS is the same as that of the AF-RIS, except that the amplifying and filtering circuit is replaced by a microstrip line. Through a combination of numerical analysis and full-wave simulation, we can obtain the gain of the AF-RIS and lossy RIS with different steering angles in comparison to a metallic plate of the same size. As shown in Fig. 2f, when the steering angle varies between 0°, 10°, 20°, and 30°, the gain of the lossy RIS and AF-RIS ranges from -8.5 to -4.9 dB and 17.2 to 21.1 dB within 2.8 to 3.2 GHz, respectively. It is observed that the AF-RIS shows an over 20 dB energy enhancement in the passband. Moreover, as we control the gain of the integrated circuit, the AF-RIS can realize a wide range of tuning of the reflection amplitude in the passband, shown in Fig. 2g. This implies that the AF-RIS can dynamically control the amplitude of the reflected EM wave, rather than simply enhancing its energy at a fixed level. Furthermore, as we can see from the curves in Fig. 2f, the gain of AF-RIS exhibits a steeper out-of-band reduction compared to the lossy RIS, showing great filtering performance. For a better illustration of the filtering ability, the E-field intensity distribution from 2.6 to 3.4 GHz is displayed in Fig. 2h. At 2.8, 3.0, and 3.2 GHz, the EM waves can be successfully reflected in the expected direction. In contrast, at 2.6 and 3.4 GHz, the excellent rejection performance of the AF-RIS is observed in the stopbands. These simulation results effectively illustrate the amplifying and filtering functionality of the proposed AF-RIS, allowing the in-band amplification and out-of-band rejection of the EM waves.

**Low hardware cost, low energy consumption, and miniaturization advantages**

When applied in wireless communication systems, the additional path loss attenuation in the relay links typically requires the increasing size of RIS arrays to ensure adequate signal strength, resulting in a high hardware cost. As analyzed in Ref. 71, the received signal power through the reflection of an RIS is proportional to the square of the RIS area. In our design, the proposed AF-RIS can provide an over 20 dB energy enhancement compared with the lossy RIS of the same size. This implies that, given the same amount of energy at the receiver, the area of the array required for AF-RIS is one-tenth that of a lossy RIS. In this case, miniaturization of the RIS array can be realized with the proposed AF-RIS, indicating a great reduction in hardware cost.

Besides, in the previous designs of A-RISs[57–62], each A-RIS element is integrated with an individual active component (amplifier or transistor). This fully-connected architecture of A-RIS results in high power consumption and hardware cost, which may mask the advantages of signal enhancement of A-RIS and hinder its practical application. To solve this problem, the power combining and dividing network is introduced in our design of the AF-RIS subarray. As shown in Fig. 2c, eight AF-RIS elements share the same filtering and amplifying circuit while they can control their phase shift independently. This sub-connected architecture significantly reduces the number of amplifiers for power saving. As revealed in Ref. 50, researchers theoretically verify that it can achieve much higher energy efficiency compared with the fully-connected structure, although at the cost of fewer degrees of freedom for beamforming design. Moreover, even when compared to a lossy RIS, the power consumption of the AF-RIS is one-third less, given the same amount of energy at the receiver. More details about the power consumption of AF-RIS can be found in Supplementary Note 4.

**Fabrication and experimental verification**

An array of 1×4 AF-RIS subarrays is fabricated utilizing printed circuit board technology and the picture of the prototype is displayed in Fig. 3a-e. A series of experiments are carried out to validate its abilities in phase-tuning, beam-steering, signal filtering, and amplifying in an indoor anechoic environment, as depicted in Fig. 3f.

Firstly, the phase tuning performance of AF-RIS is verified and the normalized phase response is measured. In the measurement, all the elements of the AF-RIS array are set at the same phase state and changed to state S0, S1, S2, and S3 sequentially. As shown in Fig. 4a, a 2-bit phase-tuning operation with the four reconfigurable coding states is realized within the passband. Owing to this great phase-tuning ability, flexible beam steering can be realized by dynamically changing the coding sequence of the AF-RIS array.

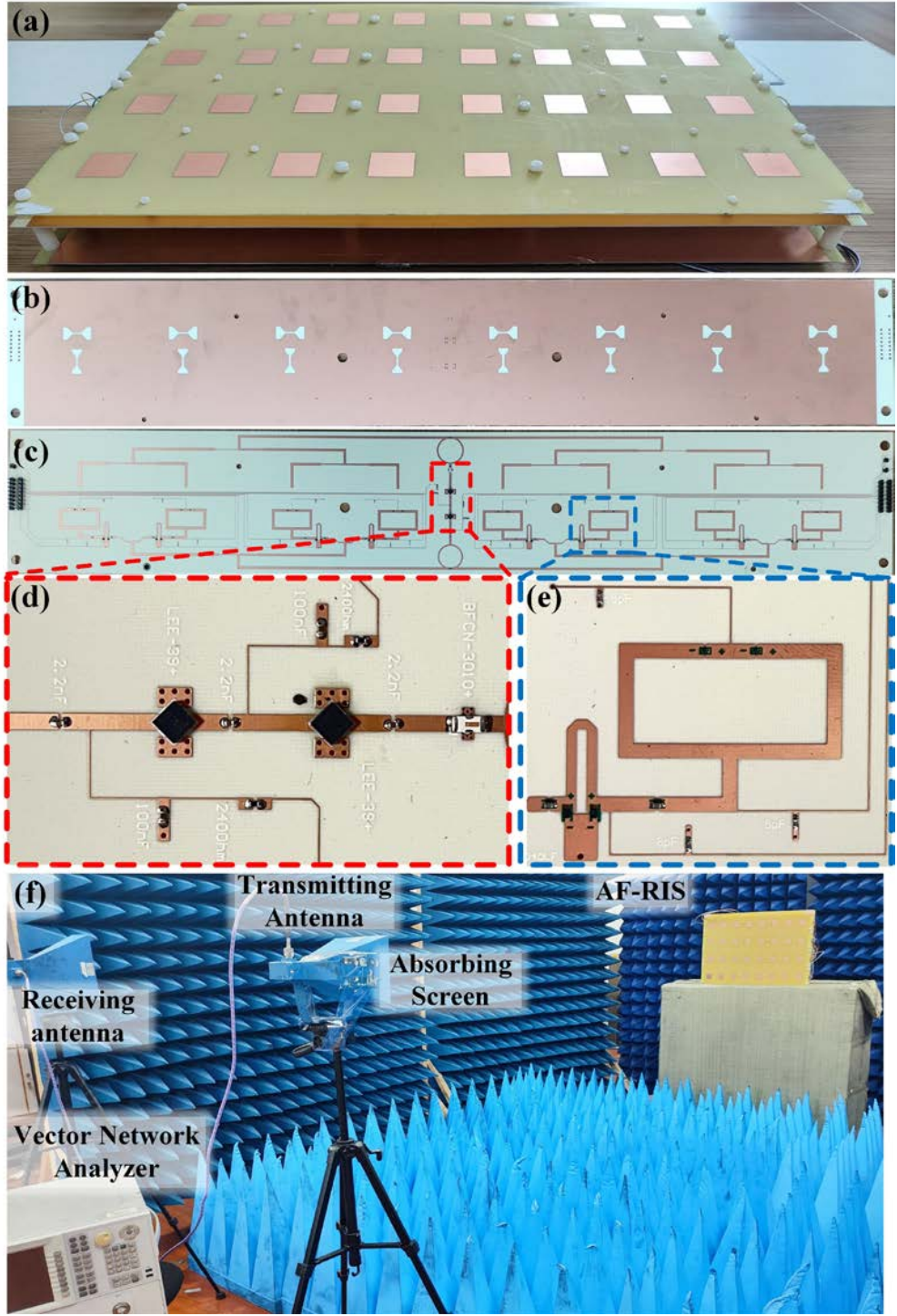

Figure 3. (a) Prototype of the fabricated AF-RIS array. (b, c) Top view of the slot plane and the microstrip network. (d, e) Detailed view of the amplifying and filtering circuits and phase-tuning structure. (g) Photograph of the simple anechoic environment.

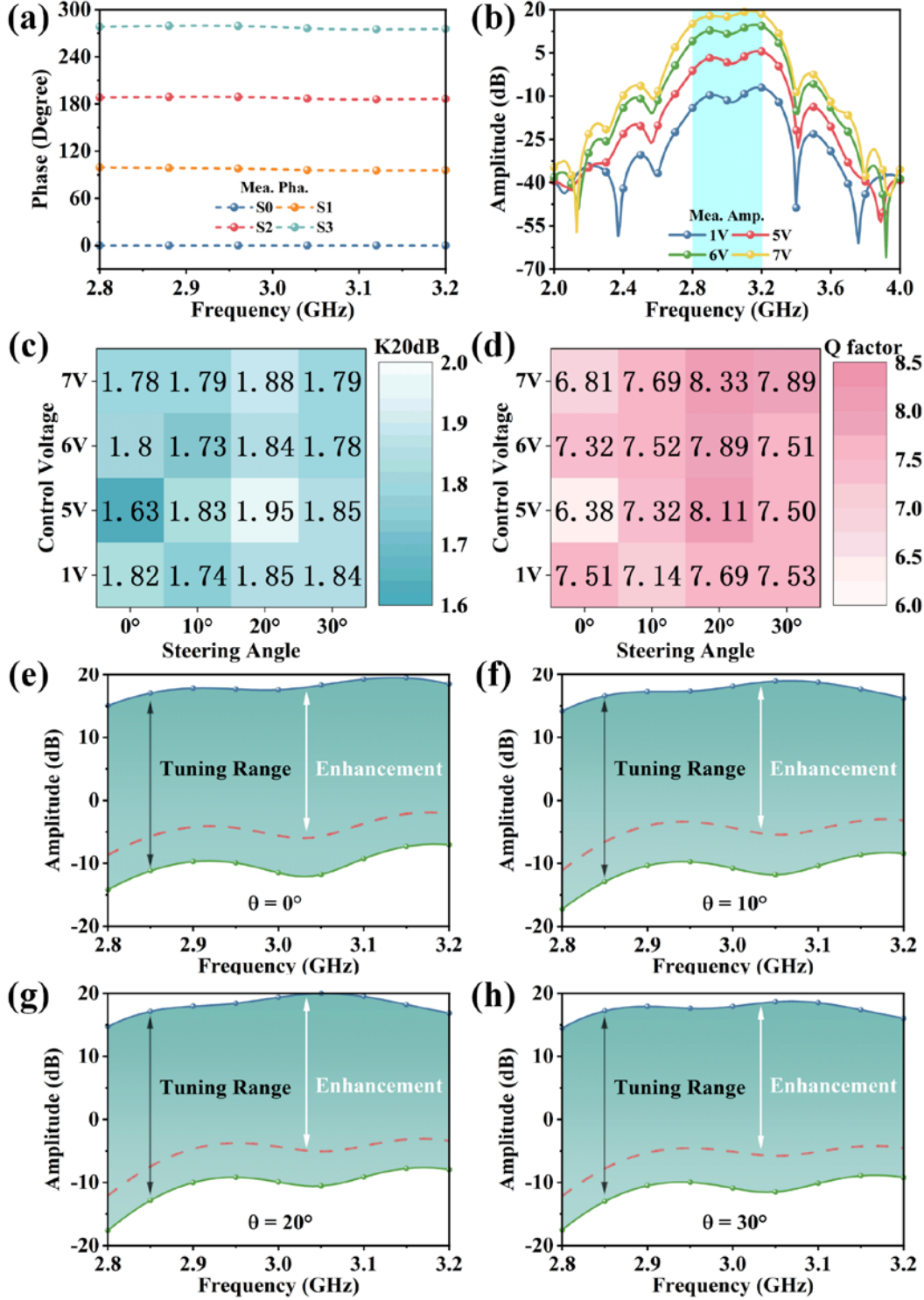

Figure 4. (a) The normalized phase response of AF-RIS with steering angle set as 0° and control voltage set as 7V. (b) The amplitude response of AF-RIS with different control voltage while the steering angle is 0°. (c, d) The rectangular coefficient K20dB and Q factor of AF-RIS with different steering angles and control voltage. (e)-(h) The amplitude responses of AF-RIS in the operating bandwidth at different steering angles, such as (a) θ = 0°, (b) θ = 10° (c) θ = 20° (d) θ = 30°.

Secondly, the filtering property of AF-RIS is validated. As we set the steering beam at 0°, the amplitude responses of AF-RIS with different control voltages of the amplifying circuit are measured and shown in Fig. 4b. It is observed that the amplitude of AF-RIS varies at different levels as we change the control voltage. However, it still exhibits stable and rapid out-of-band amplitude reduction, showing great filtering performance. For a better illustration of the filtering property, two traditional parameters (Q factor and rectangle coefficient K20dB, the definition can be found in Supplementary Note 5) are employed to quantitively measure the filtering property. Typically, the rectangular coefficient K20dB tends to exceed 1, with the optimal value of 1 indicating sharp transitions at both ends of the reflection curve, resulting in an impeccable filtering effect. Hence, an optimal RIS would boast a high Q factor and a K20dB value approaching 1. As shown in Fig. 4 c, d, those two parameters with different control voltages and steering angles are measured and listed. The closely clustered values indicate that the AF-RIS shows a stable and great filtering effect, ensuring its application in various scenarios where different beam steering angles and control voltages are required.

Finally, the in-band energy enhancement capability of the AF-RIS is displayed. The corresponding measured results are presented in Fig. 4 e-h. The upper and lower bounds are the maximum and minimum reflection amplitude of the AF-RIS, while the red dashed line is the reflection amplitude of the AF-RIS with the control voltage setting as 2V to mimic the amplitude of normal lossy RIS. The results indicate that the AF-RIS array can dynamically reflect the normal incident EM wave to the direction of 0°, 10°, 20°, and 30° with more than 20 dB energy enhancement compared with the lossy RIS. Also, at least a 25 dB amplitude tuning range is realized as we control the supply voltage from 1 V to 7 V.

Hence the fabricated AF-RIS array demonstrates dynamic phase tuning and beamforming abilities, stable frequency selectivity, commendable energy amplification, and flexible reflection amplitude tunning, which provide a solid hardware foundation for further applications in real wireless communication scenarios. More details about the amplifying and filtering performance of AF-RIS can be found in Supplementary Information Notes 5.

**Validation of AF-RIS-assisted wireless communication**
To validate the performance enhancement aroused by the AF-RIS array in real wireless communication systems, a series of experiments are conducted using a software-defined radio (SDR) platform (NI USRP-2974). As can be seen in Fig. 5, two horn antennas are connected to the SDR platform for signal transmitting and receiving. The AF-RIS array is positioned in the normal direction of the TX antenna as the wireless relay, while the RX antenna is placed at a proper receiving angle θ. A video source is converted into a bit stream, modulated by quadrature phase shift keying (QPSK) modulation scheme, transmitted to the AF-RIS array, relayed to the RX antenna via beamforming, and subsequently received by the SDR platform itself. To illustrate the performance of our AF-RIS array effectively, the energy of the transmitted signal is set at a relatively low level of -10 dBm.

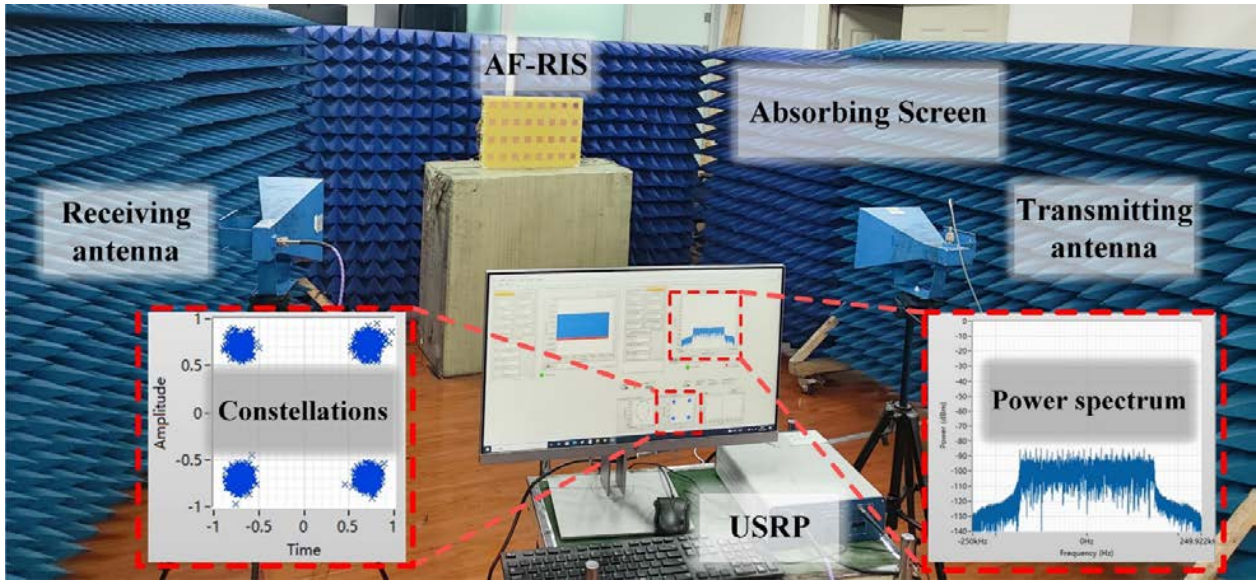

Figure 5. Photograph of the wireless communication system with AF-RIS.

Table 1 | Settings of AF-RIS and receiving antennas in the wireless communication experiments

| Cases | Theoretical beam Direction | Receiving Antenna Direction | Amplifying Mode (Y/N) | Operating Frequency |
|---|---|---|---|---|
| 1 | 0° | 30° | N | 3.00 GHz |
| 2 | 30° | 30° | N | 3.00GHz |
| 3 | 30° | 30° | Y | 3.00 GHz |
| 4 | 30° | 30° | Y | 2.65 GHz |
| 5 | 30° | 30° | Y | 3.35 GHz |

Five different cases are designed to showcase the effectiveness of the AF-RIS array, and the corresponding energy spectrums and constellation diagrams are measured and obtained (see Fig. 6). Settings of AF-RIS and directions of the Rx. antennas are summarized in Table 1. In Case 1, the carrier frequency is set at 3 GHz, and the control voltage of the amplifying circuit in the AF-RIS array is set at 2V, acting as a lossy RIS without energy amplification for fair comparison. The array is set to reflect the beam in the normal direction, not towards the Rx. antenna. As we can see in Fig. 6a, the received signal power is very low, the constellation diagram is completely cluttered, and there is no video image on the screen of the receiving terminal, which means the information transmission is completely interrupted. Then in Case 2, the control voltage of the amplifiers is unchanged while the reflect beam of the array is set to point towards the Rx. antenna. In this scenario, as we can see in Fig. 6b, the power of the received signal is enhanced, and the constellation diagram is obviously improved. This can be attributed to the gain introduced by the beamforming function of the array. However, the signal transmission still suffers a severe deterioration (see the mosaic artifacts in the received video). In this case, the energy of the transmitted signal is set at a low level as mentioned above, which means the extra gain of the beamforming function is not enough to provide adequate signal energy enhancement to improve the communication quality. So, further in-band signal energy enhancement is considered in Case 3, where the control voltage of the amplifying circuit is switched to 7 V, indicating a significant energy amplification. As shown in Fig. 6c, the signal power is clearly elevated, leading to a much more cohesive and condensed clustering of constellation points in the QPSK constellation diagram. And the video is transmitted smoothly without any disruptions, ensuring uninterrupted playback at the receiving end. It can be seen that with extra energy amplification provided by the AF-RIS array, the communication quality is highly improved compared with Cases 1 and 2. At last, the out-of-band signal rejection is also verified in our test. In Case 4 and Case 5, with other conditions remaining unchanged, the signal frequency is set at 2.65 GHz and 3.35 GHz. The corresponding measured results are illustrated in Fig. 6d-e. As shown in the energy spectrums, the energy of out-of-band signals is successfully suppressed, avoiding potential interference with the current wireless communication systems. The worsened constellation diagram and the deterioration of the video transmission also strongly suggest that such signals are rejected by the AF-RIS, indicating its great filtering performance.

To conduct a more detailed quantitative assessment of signal transmission performance, we analyze the received signal-to-noise ratio (SNR) relative to the power of the transmitted signal across Cases 1 to 3, as depicted in Fig. 6f. A noticeable trend reveals that, in Case 1 and Case 2, there is a corresponding incremental improvement in the SNR as the signal power increases, and in Case 3, the received SNR becomes relatively stable. This is because when transmitter noise is lower than receiver noise, the received SNR will increase obviously with the transmitted power.

While when the transmitter noise prevails in the overall system noise, it will not change dramatically since the AF-RIS amplifies the signal and noise from the transmitter equally. As analyzed, when the signal power is at a relatively low level, the AF-RIS can provide considerable SNR enhancement compared to the lossy RIS. Figure 6g further shows the received SNR at different carrier frequencies in Case 3 with the signal power of -10 dBm. It can be observed that the peak value of 27.1 dB is evident at the frequency of 3.0 GHz, surpassing 20.5 dB within the range of 2.8 to 3.2 GHz. Besides, beyond the 2.5 to 3.4 GHz span, the value sharply declines to below 11.7 dB, highlighting the frequency selection property of the AF-RIS.

The above discussion suggests that the AF-RIS, acting as a wireless relay, delivers enhanced signal energy, superior frequency selectivity, and improved communication quality compared to the lossy RIS. This approach may present a novel solution for extending signal coverage, mitigating spectrum pollution, and minimizing the size of the RIS array.

## Discussion

We proposed a novel AF-RIS that combines in-band signal amplification, out-of-band signal filtering, and dynamic beam-steering ability to assist wireless communication systems. By adopting the design of dual-polarized slot-coupled patches integrated with the filtering and amplifying circuit as well as the phase-tuning structure, an AF-RIS prototype is fabricated and its performance is measured in a simple anechoic chamber. The measured results show 20 dB energy enhancement in the passband and great signal reduction in the stopbands, which agrees well with the simulation results. Moreover, the 2-bit phase tuning ability enables the dynamic manipulation of the wave propagation, ensuring great filtering and amplifying performance at different angles. Thereafter, we demonstrate the amplifying, filtering, and beam steering properties of the proposed AF-RIS in the wireless communication scenario and it shows a commendable relay performance during the experiments. The AF-RIS array integrates good in-band signal amplification, out-of-band rejection, and dynamic beamforming functionalities with low hardware cost and power consumption, making us anticipate that it will be valuable to enhance the signal coverage, alleviate spectrum pollution, and reduce the size of the metasurface arrays in the 6G wireless communication system.

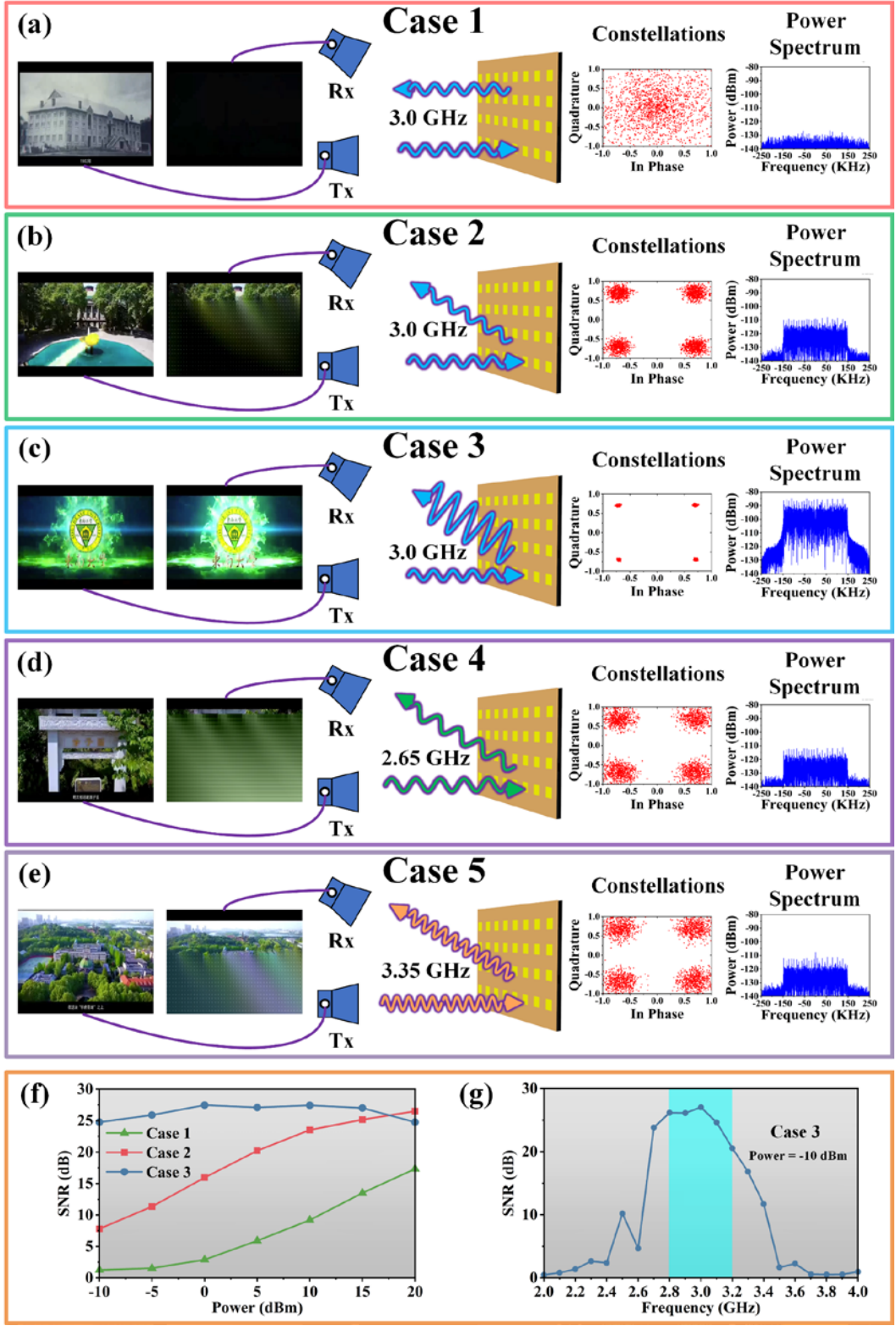

Figure 6. The wireless communication experiments to demonstrate the energy amplification and frequency filtering of the AF-RIS. (a-e) Five cases with different settings of the AF-RIS and receiving antennas. (f) The SNR spectrum in relation to the power of the transmitted signal in Case 1,2,3. (g) The SNR spectrum in Case 3 with the transmitted signal power of -10 dBm.

## Methods
**Prototype fabrication**

Three dielectric layers of AF-RIS prototype are fabricated independently using the printed circuit board technology and then assembled with the plastic bolts and nuts to secure the entire structure. As shown in Fig. 3(a), the upper patches were etched onto the FR4 substrate, while the slot plane and the microstrip network were etched onto the two sides of the Rogers RO4350B substrate separately. The ground plane was placed beneath the slot plane to eliminate back radiation. Two air layers, 7.5 mm and 30 mm thick, were sandwiched between these three layers. The fabricated AF-RIS array measures a size of 550 mm*360 mm. An expansion area was included alongside the AF-RIS array to facilitate handling and assembly, as shown in Fig.3 (b)(c). The detailed view of the amplifying and filtering circuits and phase-tuning structure is shown in Fig.3 (d)(e).

**Measurement setup**

The fabricated sample is measured in a simple anechoic environment using the free space method. As shown in Fig. 5, two horn antennas are connected to the vector network analyzer (Agilent N5245A) to transmit and receive microwave signals. The sample is surrounded by the absorbers to avoid undesired scattering. The distance between the antennas and the sample is maintained at 5.5 m during the measurement. Additionally, a DC voltage source is linked to the sample to provide the necessary supply voltage. In the measurement, the transmitting antenna remains fixed in the normal direction of the sample, while the receiving antenna is placed in various receiving directions. According to previous discussions, the incident EM wave is received, filtered, amplified, and retransmitted in orthogonal polarization through the AF-RIS array. Thus, the measurement could be divided into three steps. First, the receiving antenna is set close to the transmitting antenna to ensure the receiving angle is close to 0°. The two antennas are carefully separated to ensure high isolation between them. Co-reflection coefficients of a metallic control plate of identical size are measured to calibrate the free space path loss. Then, the receiving antenna is rotated by 90° to obtain the cross-reflection coefficients of the AF-RIS array. Finally, the reflection beam of the AIM array is switched to different directions and the receiving antenna is adjusted to the corresponding angle to measure the cross-reflection coefficients of the sample. Throughout the movement, the distance between the sample and the receiving antenna should remain unchanged to ensure the accuracy of the test.

**Acknowledgments**

This work is supported by the National Key Research and Development Program of China (2023YFB3811502), the National Science Foundation (NSFC) for Distinguished Young Scholars of China (62225108), the National Natural Science Foundation of China (62288101, 62201139), the Jiangsu Province Frontier Leading Technology Basic Research Project (BK20212002), the Jiangsu Provincial Scientific Research Center of Applied Mathematics (BK20233002), the Program of Song Shan Laboratory (Included in the management of Major Science and Technology Program of Henan Province) (221100211300-02, 221100211300-03), the 111 Project (111-2-05), the Fundamental Research Funds for the Central Universities (2242022k60003, 2242024RCB0005, 2242024K30009), and the Southeast University - China Mobile Research Institute Joint Innovation Center (R202111101112JZC02)

**Supplementary Information for**

**A wideband amplifying and filtering reconfigurable intelligent surface for wireless relay**

Lijie Wu [†][1], Qun Yan Zhou [†][1], Jun Yan Dai [†][*][1,2,3], Siran Wang [†][1,4], Junwei Zhang [1], Zhen Jie Qi [1], Hanqing Yang [1], Ruizhe Jiang [1], Zheng Xing Wang [1], Huidong Li [1], Zhen Zhang [6], Jiang Luo [7], Qiang Cheng [*][1,2,3], and Tie Jun Cui [*][1,2,5]

1. State Key Laboratory of Millimeter Waves, Southeast University, Nanjing 210096, China
2. Institute of Electromagnetic Space, Southeast University, Nanjing 210096, China
3. Frontiers Science Center for Mobile Information Communication and Security, Southeast University, Nanjing 210096, China
4. State Key Laboratory of Terahertz and Millimeter Waves, City University of Hong Kong, Hong Kong, China
5. Suzhou Laboratory, Suzhou 215000, China
6. School of Electronics and Communication Engineering, Guangzhou University, Guangzhou 510006, China
7. School of Electronics and Information, Hangzhou Dianzi University, Hangzhou, 310018, China

[†]Equally contributed to this work
[*]E-mail: junyand@seu.edu.cn; qiangcheng@seu.edu.cn; tjcui@seu.edu.cn

**Supplementary Note S1: Design and performance of the AF-RIS element**

The detailed structure of the AF-RIS element is presented in Fig. S1. It comprises four main components: the top patch, the middle slot plane, the microstrip network, and the bottom ground. The top square patch with a width of 30.87 mm is etched on the FR4 substrate with a dielectric constant $\varepsilon_r$ of 4.3 and a loss tangent tan δ of 0.025. The slot plane and the microstrip network are separated by the Rogers RO4350B substrate with a dielectric constant $\varepsilon_r$ of 3.66 and a loss tangent tan δ of 0.0037. A ground plane, etched on the FR4 substrate, is positioned beneath the microstrip network to mitigate back radiation. The primary operational principle of the AF-RIS element is as follows: The top patch receives the EM wave and couples it into the microstrip network for frequency filtering, energy amplification, and phase adjustment. Then, the corresponding EM wave is recoupled through the slot plane to the patch for reradiation.

For a better illustration of the working mechanism of AF-RIS, a detailed view of the slot plane and microstrip network is presented in Fig. S1b. As demonstrated in Fig. S1c, two hourglass-shaped slots are elaborately designed to facilitate energy coupling, enabling effective impedance matching between the patch and microstrip network over a broad frequency band. Specifically, the incident *x*-polarized EM wave is received by the top patch and coupled to the microstrip lines through the *x*-polarized slot. Then, the coupled EM wave is conveyed to the loaded filter and amplifier for signal filtering and enhancement. After that, the EM wave is guided through the phase-tuning structure, as depicted in Figs. S1d and S1e. With the control of the PIN diodes in the 0°/90° phase shifter and 0°/180° switch, the EM wave can traverse different microstrip paths to achieve accurate 2-bit phase tuning. Finally, the EM wave is recoupled through the *y*-polarized slot and reradiated by the patch in the *y*-polarization.

To verify the validity of the proposed AF-RIS element, full wave simulations were conducted with CST MWS. As shown in Fig. S2a, the insertion losses of the proposed 0°/90° phase shifter in two different states remain below 1 dB within 2.8 to 3.2 GHz, while exhibiting a phase shift ranging from 95° to 85°. The 0°/180° switch is realized based on a current inversion mechanism, wherein a 180° phase difference is achieved through a pair of oppositely arranged PIN diodes, ensuring a wide and stable phase tuning range. Thus, the AF-RIS element can achieve 2-bit phase tunning as depicted in Fig. S2b. Additionally, the S parameters of the AF-RIS without the amplifier and filter are illustrated in Fig. S2c. It is evident that the reflection coefficients of both ports ($S_{11}$ and $S_{22}$) at different phase states are maintained below -10 dB from 2.8 to 3.2 GHz, ensuring the broad operating band of AF-RIS. Furthermore, a high isolation of 30 dB is attained within the operating bandwidth (see the $S_{21}$ curve in Fig. S2c), guaranteeing the stability of the amplifier.

The simulated reflection amplitude of the AF-RIS under normal incidence at four phase states is illustrated in Fig. S2d. The reflection coefficients are denoted as R*x-y*. The left and right parts of the subscripts divided by "-" in the notations represent the incident and reflection polarization directions, respectively. For comparison, the lossy RIS is considered here, where ports 1 and 2 are directly connected. The amplitude of the lossy RIS at the four phase states ranges from -6.3 to -3.8 dB within the 2.8 to 3.2 GHz range, indicating effective energy conversion from *x*-polarized waves to *y*-polarized waves. For the normal AF-RIS element, the filter (BFCN-3010+, Mini-Circuit Ltd.) and amplifier (LEE-39+, Mini-Circuit Ltd.) are integrated between ports 1 and 2 to realize frequency selectivity and energy enhancement. As demonstrated in Fig. S2d, the reflection amplitude at each phase state exceeds 8 dB within 2.8 to 3.2 GHz, signifying much stronger energy in the reflected *y*-polarized EM waves compared to the incident *x*-polarized waves. In addition, the amplitude decreases to below -21 and -16 dB at 2.6 GHz and 3.4 GHz, respectively, leading to a great out-of-band rejection. Furthermore, it is noteworthy the AF-RIS element with the four different phase states exhibits similar performance, with only minor discrepancies in the amplitude response at each state.

In the design above, the polarization conversion ensures the high isolation between the input and output ports of the amplifier, which avoids the undesired oscillation. The integrated filter provides great bandpass response and out-of-band rejection. The elaborate design of the slot-coupled structure brings a wide operating band. Besides, the combination of 0°/90° phase shifter and 0°/180° switch provides the 2-bit phase tuning ability. Thus, the proposed AF-RIS element can achieve energy amplification, frequency selectivity, and 2-bit phase tuning in a wide band.

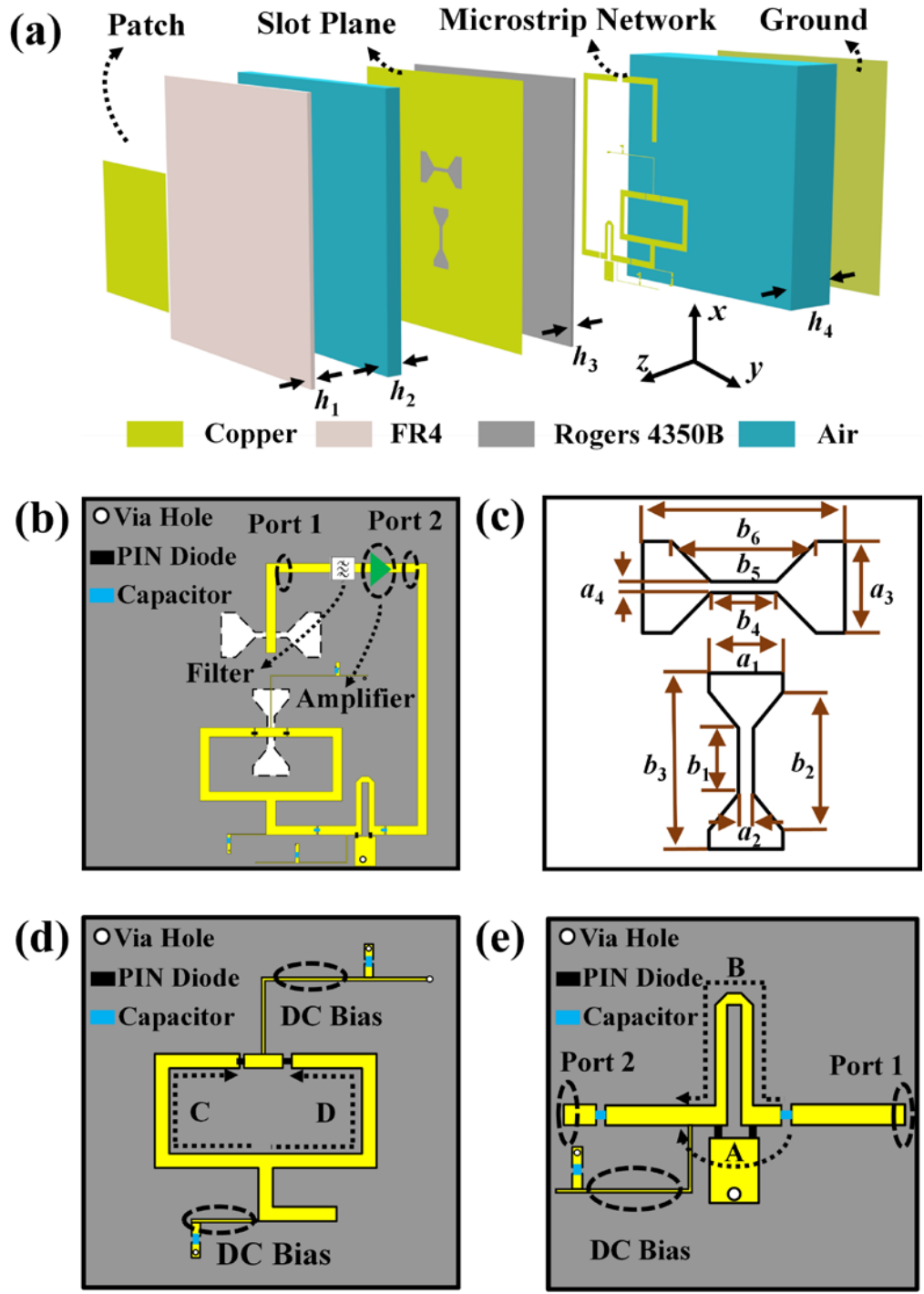

**Figure S1. The structure of the AF-RIS element.** (a) Exploded view of the proposed AF-RIS element, which $h_1 = 1$, $h_2 = 7.5$, $h_3 = 0.762$, $h_4 = 30$ (unit: mm). (b) Top view of the slot plane and the microstrip network. (c) Detailed view of the x-polarized and y-polarized slot, in which $a_1 = 6.4$, $a_2 = 1.3$, $a_3 = 8.0$, $a_4 = 0.9$, $b_1 = 5.8$, $b_2 = 12.1$, $b_3 = 15.4$, $b_4 = 5.7$, $b_5 = 12.6$, $b_6 = 17.5$ (unit: mm). (d-e) Detailed view of 0°/90° phase shifter and 0°/180° switch.

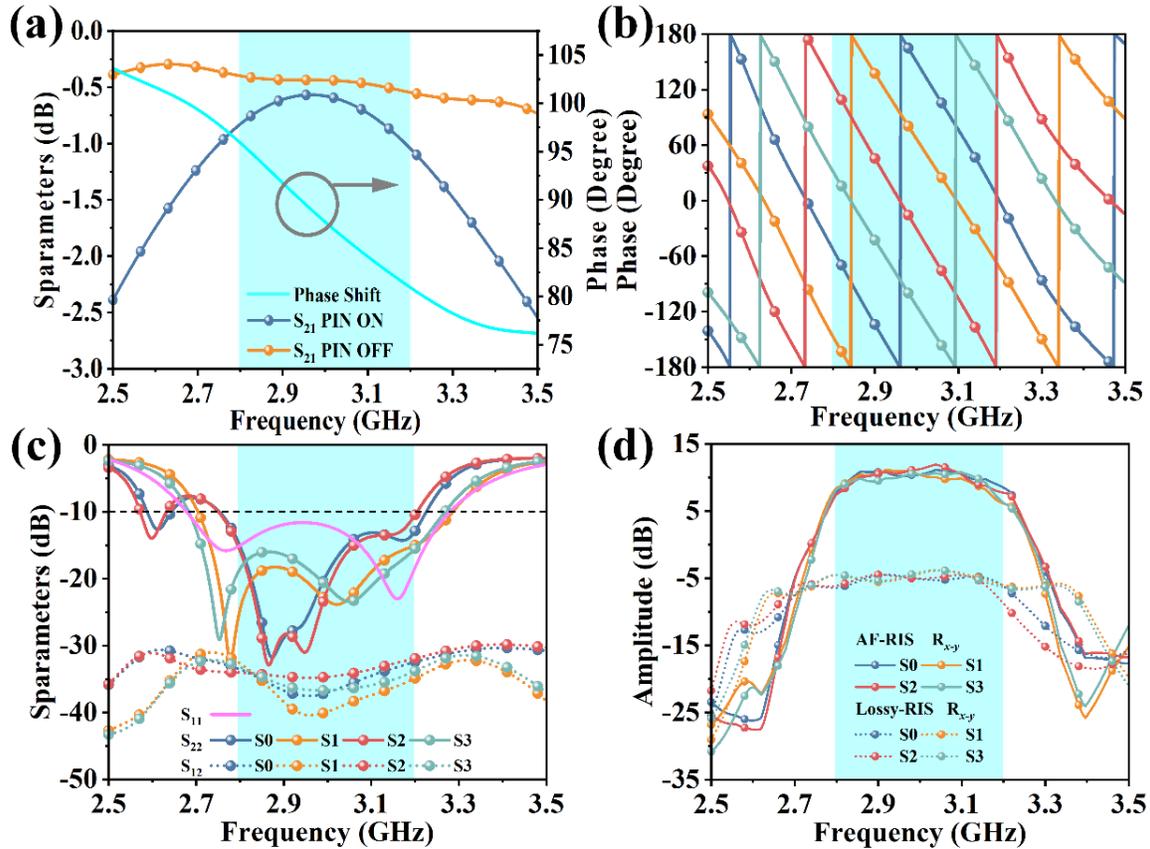

**Figure S2. Simulation results of the AF-RIS element.** (a) Simulated transmission coefficients ($S_{21}$) and phase shift of the 0°/90° phase shifter. (b) Reflection phase response of the AF-RIS element. (c) S parameters of the AF-RIS element at different phase states. (d) Reflection amplitude of the lossy RIS and AF-RIS element under normal incidence at four phase states.

**Supplementary Note S2: Design and performance of the filtering and amplifying circuit**

The BFCN-3010+ filter and LEE-39+ amplifier are integrated into the AF-RIS subarray, as shown in Fig. 3c. In our design, two kinds of filtering and amplifying circuits are considered, as shown in Figs. S3a and S3b. To prevent coupling of the DC and RF signals, capacitors $C_1$, $C_2$, and $C_3$ of 2200 pF are employed to block the DC signal on the RF path. Furthermore, capacitors $C_4$ and $C_5$ of 100 nF are utilized to bypass the RF signal on the DC path. Biasing resistors $R_2$ and $R_3$ of 243 Ω are used to establish an appropriate current for the amplifier. In the case of a single amplifier (see Fig. S3a), the LEE-39+ amplifier is cascaded with a 0 Ω resistor $R_1$, and the measured gain of the circuit is displayed in Fig. S3c. The circuit with one LEE-39+ amplifier could yield a gain of 11.2 dB to 14.1 dB within 2.8 to 3.2 GHz, which is 1.5 dB to 2 dB less than the simulation results from 2.9 GHz to 3.2 GHz. To realize more flexible control of the amplitude of the AF-RIS subarray, the circuit with two cascaded LEE-39+ amplifiers is fabricated (see Fig. S3b). To avoid oscillation, the control voltage of the first amplifier was normally set as 12 V, whereas that of the second amplifier was set to below 7 V. As shown in Fig. S3d, the measured gain of the cascaded amplifiers ranged from -1.1 dB to 26.5 dB, with the control voltage of the second amplifier adjusted from 1.0 V to 7.0 V. The design of two cascaded amplifiers is finally adopted in the structure of the AF-RIS subarray, owing to the higher amplification and more flexible gain control.

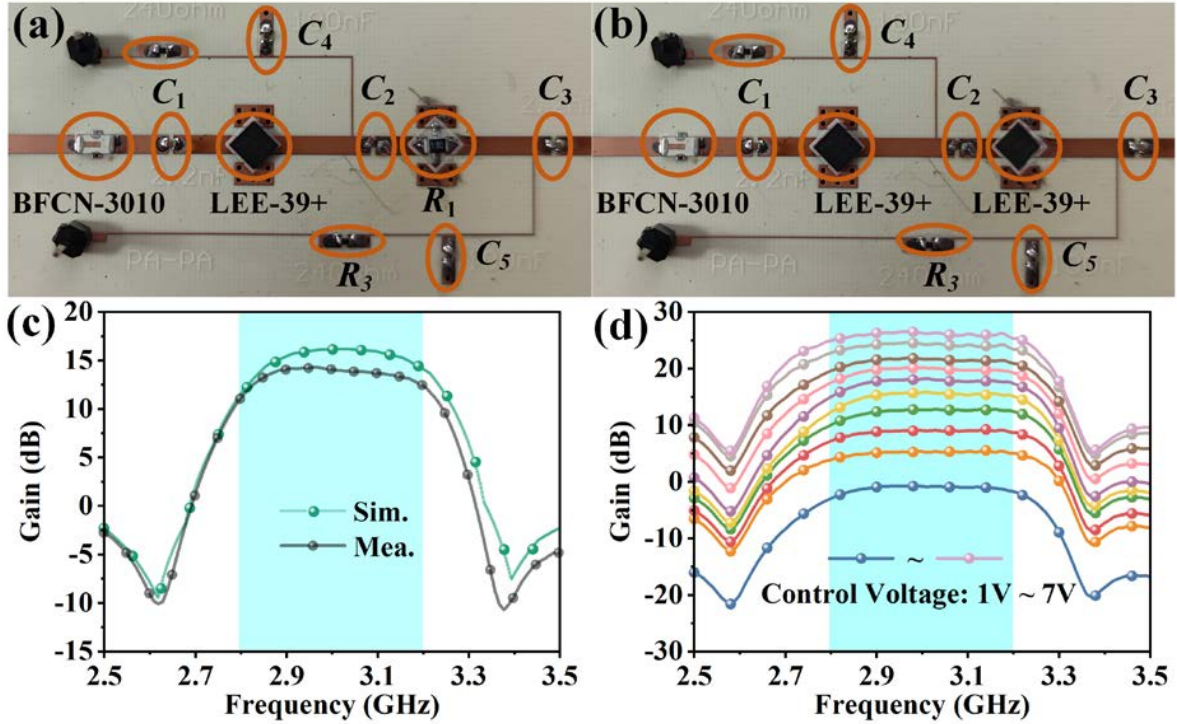

**Figure S3. Photos and gain performance of two filtering and amplifying circuits.** (a), (c) illustrate the circuit with one amplifier and gain, respectively; (b), (d) showcase the circuit with two cascaded amplifiers and gain, respectively.

**Supplementary Note S3: Power combining and dividing network design**

The power dividing and combining networks share a similar structure, as depicted in Fig. S4. The design cascades 3-level T-junctions, where the first level circuit is inspired by the Wilkson power divider, ensuring effective port isolation. As shown in the figure, the upper power combining network is an equal 1-to-8 power combiner, while the lower power dividing network is a 1-to-8 tapered power divider. The design enables the AF-RIS to share the same amplifying and filtering circuit, which reduces the usage of filters and amplifiers.

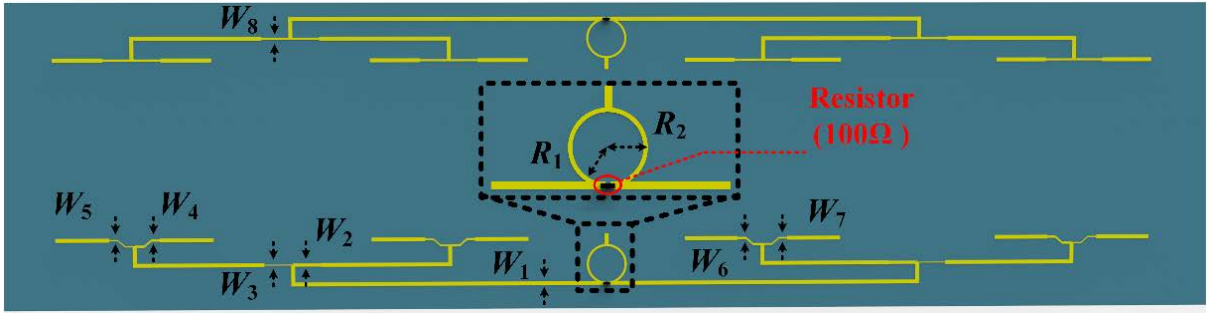

**Fig. S4. Geometric details of the power dividing and combining network.** $R_1$=8.00, $R_2$=8.90, $W_1$=1.65, $W_2$=1.35, $W_3$=0.40, $W_4$=0.99, $W_5$=0.57, $W_6$=0.94, $W_7$=0.73, $W_8$=0.89 (unit: mm).

**Supplementary Note S4: Analysis of the power consumption of AF-RIS and lossy RIS**

The proposed AF-RIS array comprises four identical AF-RIS subarrays, and each subarray consists of eight elements and one filtering and amplifying circuit. Considering the scenario of maximum power consumption, each element requires 3 PIN diodes (MACOM, MADP- 000907) operating together with a voltage of 1.33 V and a current of 30mA (details of the forward voltage and current can be found in the datasheet of the PIN diodes). Thus, the maximum power consumption for each element is 39.9 mW. For the filtering and amplifying circuit, the first amplifier operates at 12 V with a current of 35 mA, and the second amplifier operates at 7 V with a current of 15 mA, with a total power consumption of 525 mW. In this case, the power consumption of the subarray is about 844 mW.

For lossy RIS, as analyzed in the main text, it requires an area ten times larger than that of an AF-RIS to deliver the same amount of energy to the receiving end. Compared to the subarray of AF-RIS, eighty elements are needed in the lossy RIS with a total power consumption of 3192 mW. It is evident that the AF-RIS consumes less power than the lossy RIS while providing the same adequate signal power.

**Supplementary Note S5: The amplifying and filtering performance of AF-RIS**

Part of the amplitude response results of AF-RIS are shown in Figs. 4(e-h), exhibiting great signal amplification and wide-range tuning. Here, for additional clarification, the amplitude response results of AF-RIS at different directions with tunable reflection amplitude are presented in Fig. S5. The results indicate that when the control voltage is set at 7 V, the AF-RIS array can dynamically reflect the normal incident EM wave to the direction of 0°, 10°, 20°, and 30° with more than 15 dB energy enhancement, showcasing the remarkable beamforming and amplifying ability of the AF-RIS array. Furthermore, by adjusting the control voltage from 1 V to 7 V, the AF-RIS array can achieve more than 25 dB amplitude tuning of the reflection wave.

In addition, as illustrated in Figures 4c and 4d, two parameters, namely the Q factor and the rectangle coefficient K20dB, are utilized to quantitatively evaluate the filtering characteristics. In this context, a comprehensive elucidation of these two parameters is provided. Initially, the notion of BW$n$dB is introduced, denoting the bandwidth within which the reflection amplitude is reduced by less than $n$ dB from its peak value. The rectangle coefficient K20dB is determined by the ratio of BW20dB to BW3dB, whereas the Q factor is calculated as the ratio of the center frequency $f_0$ to BW3dB. Fig. S6 illustrates the BW3dB and BW20dB values for AF-RIS under various steering angles and control voltages, with closely matched values indicating consistent filtering performance stability.

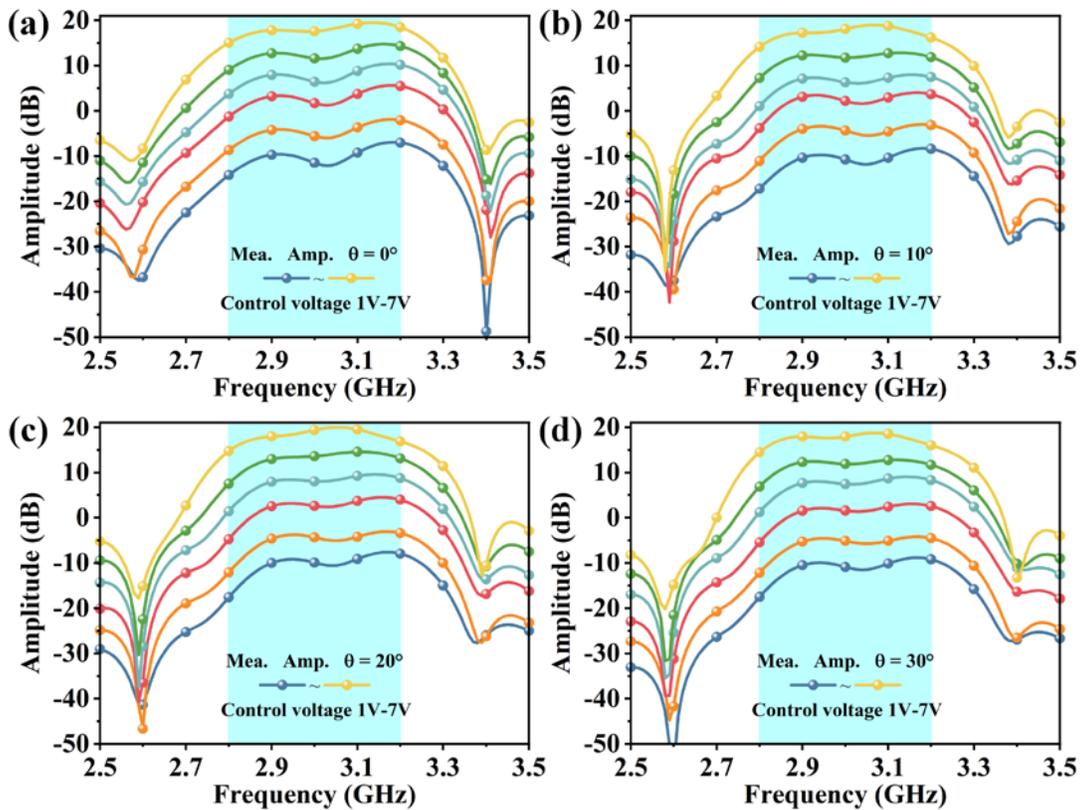

**Figure S5. Measured reflection amplitudes of the AF-RIS array across a control voltage range of 1 V to 7 V at various steering angles.** (a) Steering angle: 0°; (b) Steering angle: 10°; (c) Steering angle: 20°; (d) Steering angle: 30°.

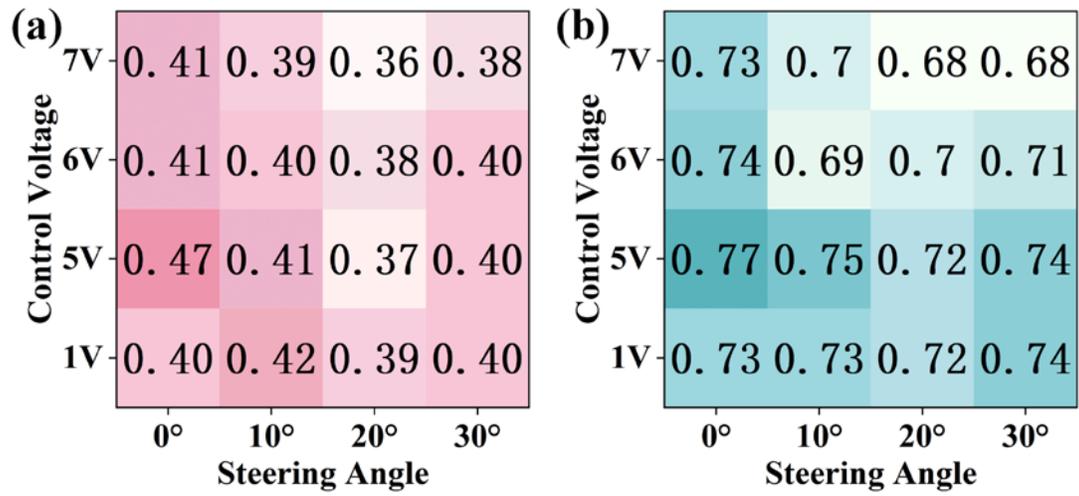

**Figure S6. The BW3dB and BW20dB values of AF-RIS with different steering angles and control voltages. (Unit: GHz)** (a) BW3dB; (b) BW20dB.